\documentclass{article}

\def\abstract#1{\vskip 7mm 
        \begin{center}{\large Abstract}\par \smallskip
                \begin{minipage}[c]{12cm}
                        \small #1
                \end{minipage}
        \end{center}
}
\def\title#1{\begin{center}{\Large\bf #1}\end{center}}
\def\author#1{\vskip 5mm \begin{center}{#1}\end{center}}
\def\address#1{\begin{center}{\it #1}\end{center}}

\makeatletter
\@ifundefined{lesssim}{}{}
\@ifundefined{gtrsim}{}{}
\def\vereq#1#2{\lower3pt\vbox{\baselineskip1.5pt \lineskip1.5pt
\ialign{$\m@th#1\hfill##\hfil$\crcr#2\crcr\sim\crcr}}}
\makeatother
\begin{document}

\begin{flushright}

\parbox{3.2cm}{
{YITP-06-02 \hfill\\
{\tt hep-th/0601100}}
 }
\end{flushright}


\title{%
Moduli Instability in Warped Compactification
}
\author{%
  Hideo Kodama\footnote{E-mail:kodama@yukawa.kyoto-u.ac.jp}
  and 
  Kunihito Uzawa\footnote{E-mail:uzawa@yukawa.kyoto-u.ac.jp}}
\address{%
Yukawa Institute for Theoretical Physics,
  Kyoto University,
  Kyoto 606-8502, Japan
}

\abstract{
We derive four-dimensional effective theories for warped 
compactification of the ten-dimensional IIB supergravity. 
We show that these effective theories allow a much wider 
class of solutions than the original higher-dimensional 
theories. This result indicates that the effective 
four-dimensional theories should be used with caution, 
if one regards the higher-dimensional theories more fundamental.
}

\section{Introduction}

Recently, a new class of dynamical solutions describing a 
size-modulus instability in the ten-dimensional type IIB 
supergravity model have been discovered by Gibbons et al. 
\cite{Gibbons:2005rt} and the authors \cite{Kodama:2005fz}. These 
solutions can be always obtained by replacing the 
constant modulus $h_0$ in the warp factor 
$h=h_0+h_1(y)$ for supersymmetric solutions by a linear function 
$h_0(x)$ of the four-dimensional coordinates $x^\mu$.
Such extensions exist for many of the well-known solutions 
compactified with flux on a conifold, resolved conifold, deformed 
conifold and compact Calabi-Yau manifold  \cite{Kodama:2005fz}. 

In most of the literature, the dynamics of the internal space, 
namely the moduli, in a higher-dimensional theory is investigated by 
utilising a four-dimensional effective theory. In particular, 
effective four-dimensional theories are used in essential ways in 
recent important work on the moduli stabilisation problem and the 
cosmological constant/inflation problem in the IIB sugra 
framework \cite{Kachru:2003aw}. 
Hence, it is desirable to find the relation between the
 above dynamical solutions in the higher-dimensional 
theories and solutions in the effective four-dimensional theory. 

In the conventional approach where the non-trivial warp factor does 
not exist or is neglected, an effective four-dimensional theory is 
derived from the original theory  assuming the ``product-type'' 
ansatz for field variables. This 
ansatz requires that each basic field of the theory is expressed as 
the sum of terms of the form $f(x)\omega(y)$, where $f(x)$ is an 
unknown function of the four-dimensional coordinates $x^\mu$, and 
$\omega(y)$ is a known harmonic tensor on the internal space. 
Further, it is assumed that the higher-dimensional metric takes the 
form $ds^2=ds^2({\rm X}_4)+ h_0^{\beta}(x)ds^2({\rm Y})$, where 
$ds^2({\rm X}_4)=g_{\mu\nu}(x)dx^\mu dx^\nu $ is an unknown four-dimensional 
metric, $h_0(x)$ is the size modulus for the internal space 
depending only on the $x$-coordinates, and $ds^2({\rm Y})=\gamma_{pq}dy^p 
dy^q$ is a (Calabi-Yau) metric of the internal space that depends on 
the $x$-coordinates only through moduli parameters. Under this 
ansatz, the four-dimensional effective action is obtained by 
integrating out the known dependence on $y^p$ in the 
higher-dimensional action. 

The dynamical solutions in the warped compactification mentioned at 
the beginning, however, do not satisfy this ansatz. Hence, in order 
to incorporate such solutions to the effective theory, we have to 
modify the ansatz. Taking account of the structure of the 
supersymmetric solution, the most natural modification of the ansatz 
is to introduce the non-trivial warp factor $h$ into the metric as 
$ds^2=h^{\alpha} ds^2({\rm X}_4)+ h^{\beta} ds^2({\rm Y})$ and assume that $h$ 
depends on the four-dimensional coordinates $x^\mu$ only through the 
modulus parameter of the supersymmetric solution as in the case of 
the internal moduli degrees of freedom. This leads to the form 
$h=h_0(x)+h_1(y)$ for the IIB models, which is consistent with the 
structure of the dynamical solutions in the ten-dimensional theory.

In the present work, starting from this modified ansatz, we study 
the dynamics of the four-dimensional effective theory and its 
relation to the original higher-dimensional theory for warped 
compactification of the ten-dimensional type IIB supergravity 
For simplicity, we 
assume that the moduli parameters other than the size parameter are 
frozen. 

\section{Ten-dimensional solutions}
In our previous work \cite{Kodama:2005fz}, we derived 
a general dynamical solution for warped compactification 
with fluxes in the ten-dimensional type IIB supergravity. 
In that work, we imposed $d\ast(B_2\wedge H_3)=0$, which 
led to a slightly strong constraint on the free data for 
the solution, especially in the case of a compact 
internal space. Afterward, we have noticed that this 
condition is not necessary to solve the field equations, 
and without that condition, we can find a more general 
class of solutions. Because we take this class as the 
starting point of our argument, we first briefly 
explain how to get a general solution without that condition. 
We omit the details of calculations because they are 
essentially contained in our previous paper \cite{Kodama:2005fz}.

We assume that the ten-dimensional spacetime metric takes the form
\begin{equation}
ds^2({\rm X}_{10})=h^{-1/2}(x,y) ds^2({\rm X}_4)
  +h^{1/2}(x,y) ds^2({\rm Y}_6),
\label{IIB:metric:general}
\end{equation}
where ${\rm X}_4$ is the four-dimensional spacetime with 
coordinates $x^\mu$, and ${\rm Y}_6$ is the six-dimensional 
internal space.
We further require that the dilaton 
and the form fields satisfy the following conditions:
\begin{eqnarray}
& &\tau\equiv C_0+i\, e^{-\Phi}=ig_s^{-1}(={\rm const})\,,
      \qquad G_3\equiv ig_s^{-1}\,H_3-F_3
  =\frac{1}{3!}\,G_{pqr}(y)\,dy^p\wedge dy^q \wedge dy^r\,,
      \nonumber\\
& &\ast_{\rm Y}\,G_3= \epsilon i\,G_3\quad(\epsilon=\pm1)\,,
      \qquad
\tilde F_5=dC_4+B_2\wedge F_3
   =(1\pm\ast){V_p dy^p} \wedge \Omega({\rm X}_4),
  \label{IIB:gauge-assumption}
\end{eqnarray}
where $g_s$ is a constant representing the string 
coupling constant, and $\ast$ and $\ast_{\rm Y}$ are 
the Hodge duals with respect to the ten-dimensional 
metric $ds^2({\rm X}_{10})$ and the six-dimensional metric 
$ds^2({\rm Y}_6)$, respectively. 

Under these assumptions, we obtain the following expressions 
from the field equations for form fields 
\cite{Kodama:2005cz}:
\begin{equation}
\tilde F_5= \pm\epsilon(1\pm\ast) d(h^{-1}) \wedge \Omega({\rm X}_4),
 \qquad
\triangle_{\rm Y} h = -\frac{g_s}{2}(G_3\cdot \bar G_3)_{\rm Y}.
\label{hbyG3}
\end{equation}
Next, by using the equations 
(\ref{IIB:gauge-assumption}) and (\ref{hbyG3}), the ten-dimensional 
Einstein equations can be written 
\begin{equation}
h R_{\mu\nu}({\rm X}_4)- D_{\mu}D_{\nu}h
 +\frac{1}{4}g_{\mu\nu}({\rm X}_4)\triangle_{\rm X}h=0, \qquad
\partial_{\mu}\partial_{p}h=0,\qquad
R_{pq}({\rm Y}_6)-\frac{1}{4}g_{pq}({\rm Y}_6) \triangle_{\rm X}h=0.
\label{IIB:W-Einstein}
\end{equation}
{}From the second of these equations, we immediately 
see that the warp factor $h$ can be expressed as
$h(x,y)=h_0(x)+h_1(y)$.
Further, if we require that $d_y h \ne 0$, 
the rest of the equations can be reduced to
\begin{equation}
R_{\mu\nu}({\rm X}_4)=0, \qquad 
D_{\mu}D_{\nu}h_0= \lambda g_{\mu\nu}({\rm X}_4), \qquad
R_{pq}({\rm Y}_6)=\lambda g_{pq}({\rm Y}_6). 
  \label{IIB:sol}
\end{equation}

Thus, we have found that the most general solutions 
satisfying the conditions (\ref{IIB:metric:general}) 
and (\ref{IIB:gauge-assumption}) are specified by a 
Ricci flat spacetime ${\rm X}_4$, an Einstein space ${\rm Y}_6$, 
a closed imaginary-self-dual (ISD) 3-form $G_3$ on ${\rm Y}_6$, 
and the function 
$h(x,y)$ that is the sum of $h_0(x)$ satisfying 
the second of the equations (\ref{IIB:sol}) 
and $h_1(x)$ satisfying the second of the equations (\ref{hbyG3}). 
The additional constraint on $G_3$, 
$d_y[h^{-2}(B_2\cdot dB_2)_{\rm Y}]=0$, in Ref. \cite{Kodama:2005fz} 
does not appear. Further, closed ISD 3-forms on ${\rm Y}_6$ 
are in one-to-one correspondence with real harmonic 3-forms on 
${\rm Y}_6$.  Hence, this class of dynamical solutions exist 
even for a generic compact Calabi-Yau internal space, 
if we allow $h_1(y)$ to be a singular function. This 
singular feature of $h$ in the compact case with flux 
arises because $h$ is a solution to the Poisson equation 
(\ref{hbyG3}) and has nothing to do with the dynamical 
nature of the solution. It is shared by the other flux 
compactification models.

Here, note that the Ricci flatness of ${\rm X}_4$ is required 
from the Einstein equations. This should be contrasted 
with no-warp case \cite{Kodama:2005cz}. This point is quite 
important in the effective theory issue, as we see soon. 
Moreover, one can show that the Ricci 
flatness of ${\rm X}_4$ and the second in the equations
(\ref{IIB:sol}) are consistent 
only when ${\rm X}_4$ is locally flat if $(Dh_0)^2\not=0$.  

\section{Four-dimensional effective theory} 
Now we study the four-dimensional effective theory that 
incorporates the dynamical solutions obtained in the 
previous section. For simplicity, we do not consider 
the internal moduli degrees of freedom of the metric of 
${\rm Y}_6$ or of the solution $h_1(y)$ in the present work. 
Then, in its $x$-independent subclass with $\lambda=0$, 
we have only one free parameter $h_0$. When we rescale  
$ds^2({\rm Y}_6)$ by a constant $\ell$ as $\ell^2 ds^2({\rm Y}_6) 
\rightarrow ds^2({\rm Y}_6)$, we have to rescale $h$ as $h/\ell^4 
\rightarrow h$. We can easily see that the corresponding rescaled 
$h_1$ satisfies (\ref{hbyG3}) 
again with the same $G_3$ as 
that before the rescaling. We can also confirm that the D3 
brane charges associated with the 5-form flux do not change 
by this scaling. In contrast, $h_0$ changes its value by 
this rescaling. Therefore, $h_0$ represents the size 
modulus of the Calabi-Yau space ${\rm Y}_6$. 

On the basis of this observation, we construct the four-dimensional 
effective theory for the class of ten-dimensional 
configurations specified as follows. First, we assume 
that ${\rm X}_{10}$ has the metric 
\begin{equation}
ds^2({\rm X}_{10}) = h^{-1/2}(x,y)\,ds^2({\rm X}_{4}) 
  +h^{1/2}(x,y)\,ds^2({\rm Y}_{6}),
   \label{IIB:metric}
\end{equation}
where $h=h_0(x)+h_1(y)$ and $ds^2({\rm Y}_{6})$ is a fixed 
Einstein metric on ${\rm Y}_{6}$ satisfying the third of the 
equations (\ref{IIB:sol}), 
while $ds^2({\rm X}_{4})$ is an arbitrary metric on ${\rm X}_{4}$. 
Further, we assume that the dilaton is frozen as in 
(\ref{IIB:gauge-assumption}), $G_3$ is given by a fixed 
closed ISD 3-form on ${\rm Y}_{6}$, $h_1(y)$ is a fixed solution 
to the second of the equations (\ref{hbyG3}), 
and $\tilde F_5$ is given by the first of  
the equations (\ref{hbyG3}). 
Hence, the metric of ${\rm X}_{4}$ and the function $h_0$ on it 
are the only dynamical variables in the effective theory.

The four-dimensional effective action for these variables 
can be obtained by evaluating the ten-dimensional action 
of the IIB theory
\begin{eqnarray}
S_{\rm IIB} = \frac{1}{2\tilde\kappa^2}\int_{{\rm X}_{10}}
  d\Omega({\rm X}_{10})
 \left[R({\rm X}_{10})-\frac{\nabla_M \bar\tau \nabla^M \tau}
 {2({\rm Im}\tau)^2}
   -\frac{G_3\cdot\bar G_3}{2\,{\rm Im}\tau}-\frac{1}{4}\tilde F_5^2\right]
\pm \frac{i}{8\tilde \kappa^2}\int_{{\rm X}_{10}} 
   \frac{C_4\wedge G_3\wedge \bar G_3}{{\rm Im}\tau},
\label{IIB:action}
\end{eqnarray}
for the class of configurations specified above. In general, 
there is a subtlety concerning the action of the type IIB 
supergravity, because the correct field equations can be 
obtained by imposing the self-duality condition 
$\ast\tilde{F}_5=\pm \tilde{F}_5$ after taking variation of the 
action in general. In the present case, however, since 
we are only considering configurations (\ref{hbyG3}) 
satisfying the self-duality condition, this problem 
does not affect our argument. We can obtain the "correct" 
effective action by simply inserting (\ref{hbyG3}) into 
the above ten-dimensional action. 

Inserting the expressions (\ref{hbyG3}) and the third of the equations 
(\ref{IIB:sol}) into (\ref{IIB:action}), we get \cite{Kodama:2005cz}
\begin{eqnarray}
S_{\rm IIB}=\frac{1}{2\kappa^2} \int_{{\rm X}_4} d\Omega({\rm X}_4) 
   \left[H R({\rm X}_4) +6\lambda \right],
   \label{eq:4d-action}
\end{eqnarray}
where we have dropped the surface term coming from 
$\triangle_{\rm X} h_0$ and neglected the boundary term 
in the Chern-Simons term, $\kappa=(V_6)^{-1/2}\tilde{\kappa}$,
and $H(x)$ is defined by 
\begin{equation}
H(x)=h_0(x)+ c;\quad c:=V_6^{-1}\int_{{\rm Y}_6}d\Omega({\rm Y}_6)h_1.
\end{equation}
This effective action has the same form as the case of vanishing
flux \cite{Kodama:2005cz}. Hence, it gives the 
four-dimensional field equations of the same form
 as in the no-flux case:
\begin{eqnarray}
R_{\mu\nu}({\rm X}_4)=H^{-1} 
   \left[D_{\mu} D_{\nu} H - \lambda g_{\mu\nu}({\rm X}_4)\right],
 \qquad
\triangle_{\rm X} H= 4\lambda.
\label{eq:4d-Eequation}
\end{eqnarray}
If the four-dimensional spacetime is Ricci flat, these 
equations reproduce the correct equation for $h_0(x)=H-c$ 
obtained from the ten-dimensional theory \cite{Kodama:2005cz}. 
However, the Ricci flatness of ${\rm X}_4$ is not 
required in the effective theory unlike in the 
ten-dimensional theory. Hence, the class of solutions 
allowed in the four-dimensional effective theory is much 
larger than the original ten-dimensional theory. 

In particular, the effective theory has a modular invariance 
similar to that found in the no-flux Calabi-Yau 
case with $\lambda=0$. In fact, by the conformal 
transformation $ds^2({\rm X}_4)=H^{-1}ds^2(\bar{{\rm X}}_4)$, 
(\ref{eq:4d-action}) is expressed in terms of the 
variables in the Einstein frame as 
\begin{eqnarray}
S_{\rm IIB}=\frac{1}{2\kappa^2} \int_{\bar{\rm X}_4} 
 d\Omega(\bar{{\rm X}}_4)
 \left[R(\bar{{\rm X}}_4) - \frac{3}{2} 
 (\bar{D}\ln H)^2 +6\lambda H^{-2}\right],
   \label{eq:E4d-action}
\end{eqnarray}
where $R(\bar{{\rm X}}_4)$ and $\bar{D}_{\mu}$ are  the scalar 
curvature and the covariant derivative with respect to 
the metric $ds^2(\bar{{\rm X}}_4)$. 
The corresponding four-dimensional Einstein equations 
in the Einstein frame and the field equation for $H$ 
are given by 
\begin{eqnarray}
R_{\mu\nu}(\bar{{\rm X}}_4)=\frac{3}{2}\bar{D}_{\mu}\ln H\, 
   \bar{D}_{\nu}\ln H - 3\lambda H^{-2}g_{\mu\nu}(\bar{{\rm X}}_4),
 \qquad
\triangle_{\bar{\rm X}}\ln H= 4\lambda H^{-2},
\end{eqnarray}
where $\triangle_{\bar{\rm X}}$ is the Laplacian with 
respect to the metric $ds^2(\bar{{\rm X}}_4)$. 
It is clear that for $\lambda=0$, this action and the 
equations of motion are invariant under the transformation 
$H\rightarrow k/H$, where $k$ is an arbitrary positive 
constant. 

This transformation corresponds to the following 
transformation in the original ten-dimensional metric. 
Let us denote the new metric of ${\rm X}_4$ and the function 
$h$ obtained by this transformation by $ds'{}^2({\rm X}_4)$ 
and $h'$, respectively. Then, since the transformation 
preserves the four-dimensional metric in the Einstein 
frame, $ds'{}^2({\rm X}_4)$ is related to $ds^2({\rm X}_4)$ as 
$ds'{}^2({\rm X}_4)=(H^2/k) ds^2({\rm X}_4)$. In the meanwhile, 
from $H'=k/H= h_0' + c$, $h'$ is expressed in terms of 
the original $h_0$ as
\begin{equation}
h'= \frac{k}{h_0(x)+c}- c + h_1(y).
\end{equation}
The corresponding ten-dimensional metric is written
\begin{equation}
ds^2= k^{-1}H^2 (h')^{-1/2} ds^2({\rm X}_4)+ (h')^{1/2}ds^2({\rm Y}_6).
\end{equation}
It is clear that this metric and $h'$ do not satisfy the 
original ten-dimensional field equations. Hence, the 
modular-type invariance of the four-dimensional effective 
theory is not the invariance of the original ten-dimensional 
theory.

\section{Summary}
In the present work, we have derived four-dimensional 
effective theories for the spacetime metric and the size 
modulus of the internal space for warped compactification 
with flux in the ten-dimensional type IIB supergravity. 
The basic idea was to consider field configurations 
in higher dimensions that are obtained by replacing the 
constant size modulus in supersymmetric solutions for 
warped compactifications, by a field on the four-dimensional 
spacetime. The effective action for this moduli field and 
the four-dimensional metric has been determined by 
evaluating the higher-dimensional action for such 
configurations. In all cases, the dynamical solutions 
in the ten-dimensional theories found by 
Gibbons et al. \cite{Gibbons:2005rt}, Kodama and 
Uzawa \cite{Kodama:2005fz}
were reproduced in the four-dimensional effective theories 
\cite{Kodama:2005cz}. 

In addition to this, we have found that these 
four-dimensional effective theories have some unexpected 
features. First, the effective actions of both theories 
are exactly identical to the four-dimensional effective 
action for direct-product type compactifications with no 
flux in ten-dimensional supergravities. In particular, 
the corresponding effective theory has a kind of modular 
invariance with respect to the size modulus field in the 
Einstein frame. This implies that if there is a solution 
in which the internal space expands with the cosmic expansion, 
there is always a conjugate solution in which the internal 
space shrinks with the cosmic expansion. 

Second, the four-dimensional effective theory for warped 
compactification allows solutions that cannot be obtained 
from solutions in the original higher-dimensional theories. 
The modular invariance in the four-dimensional theory 
mentioned above is not respected in the original 
higher-dimensional theory either. 
The same results hold for the heterotic M-theory\cite{Kodama:2005cz}.
This situation should be 
contrasted with the no-warp case in which the four-dimensional 
effective theory and the original higher-dimensional theory 
are equivalent under the product-type ansatz for the metric 
structure. This result implies that we have to be careful 
when we use a four-dimensional effective theory to analyse 
the moduli stabilisation problem and the cosmological 
problems in the framework of warped compactification of 
supergravity or M-theory.



\begin{thebibliography}{99}
\bibitem{Gibbons:2005rt}
  G.~W.~Gibbons, H.~L\"{u} and C.~N.~Pope,
  Phys.\ Rev.\ Lett.\  {\bf 94} (2005) 131602
  [arXiv:hep-th/0501117].

\bibitem{Kodama:2005fz}
  H.~Kodama and K.~Uzawa,
  JHEP {\bf 0507} (2005) 061 
  [arXiv:hep-th/0504193].

\bibitem{Kachru:2003aw}
  S.~Kachru, R.~Kallosh, A.~Linde and S.~P.~Trivedi,
  Phys.\ Rev.\ D {\bf 68} (2003) 046005
  [arXiv:hep-th/0301240].

\bibitem{Kodama:2005cz}
H.~Kodama and K.~Uzawa,
arXiv:hep-th/0512104.
\end{thebibliography}
\end{document}